\documentclass[12pt]{iopart}

\begin{document}
\def\lsim{\raisebox{-.4ex}{$\stackrel{<}{\scriptstyle \sim}$\,}}
\def\gsim{\raisebox{-.4ex}{$\stackrel{>}{\scriptstyle \sim}$\,}}
\newcommand{\Od}{{\cal O}}
\title{Moving dark energy and the CMB dipole}

\author{A.~L.~Maroto}

\address{Departamento de  F\'{\i}sica Te\'orica,
 Universidad Complutense de
  Madrid, 28040 Madrid, Spain}
\ead{maroto@fis.ucm.es}
\begin{abstract}
We explore the possibility that the rest frames of CMB, matter and
dark energy differ one from another, i.e. they do not converge 
on very large scales. In such a case, the usual
interpretation of the CMB dipole as being due to the relative
motion of the observer with respect to the CMB rest frame
is not appropriate. Instead, we find that the measured dipole
is due to the observer motion relative to the cosmic {\it center of mass}
rest frame. This means, in particular, that even an observer at
rest with respect to the CMB radiation could measure a non-vanishing
dipole anisotropy, provided dark energy is moving with
respect to the CMB. We also consider the consequences of moving
dark energy 
for the determination of cosmic bulk flows.

\end{abstract}

\maketitle

\section{Introduction}

Standard cosmology assumes homogeneity and isotropy of the Universe
on very large scales. The presence of
density perturbations implies that when averaged over small volumes, 
matter can have a non-vanishing streaming velocity with respect 
to the CMB radiation, the amplitude of such motions depending on 
the actual power spectrum of density fluctuations. However, 
as we take larger and larger averaging volumes, 
convergence of both reference frames is expected according to the
Cosmological Principle. In other
words, matter and radiation should share a common rest frame. 
However, the observational situation is far from clear. Recent 
large-scale peculiar velocity surveys have measured the dipole
of the peculiar velocity field on different scales, trying to determine
the volume size at which the streaming motion vanishes. Although there
is evidence of convergence on very large scales   
$\gsim 100 h^{-1}$ Mpc in some works 
\cite{Riess},  non-vanishing 
bulk flows with amplitudes $\gsim 600$ km s$^{-1}$ with respect to the CMB 
have also been measured in other surveys on those distance scales 
\cite{LP,Koc,Hudson}, 
although the results do not agree in the direction of the motion, and they 
have been argued to be affected by systematics \cite{Hudsonp}. The 
possibility that motions with such a large amplitude could be 
accomodated within
the standard model of structure formation was
studied in \cite{Strauss} and more recently in \cite{Hudsonp}.

On the other hand, the dominant contribution to the CMB dipole anisotro\-py 
is usually attributed to
the Doppler effect due to the observer motion with respect to the last
scattering surface \cite{Peebles,Padma,Kogut}. Substracting 
the contribution from the 
solar system motion relative to the Milky Way, and the
Milky Way relative to the  Local Group (LG) \cite{MW},
 the measurement of the CMB dipole 
has been used 
to obtain the relative velocity of the LG with respect to the CMB,
which according to COBE \cite{Kogut} is $627\pm 22$ km s$^{-1}$, towards Galactic
coordinates $l=276\pm 3^\circ$,
$b=30\pm 3^\circ$. This result seems to be compatible with the direct
determination of the LG velocity with respect to the rest frame defined by 
certain SNIa host galaxies \cite{Riess2}. It is however inconsistent 
with the  
result in \cite{LP}, according to which the LG 
is moving at $561 \pm 284$ km s$^{-1}$
towards $l=220^\circ$,
$b=-28^\circ$ with total angular error of $\pm \,27^\circ$, 
with respect to the frame defined by the 119 Abell clusters contained
within a $150 h^{-1}$ Mpc distance (see also \cite{Koc}). 

Some alternatives to 
the standard $\Lambda$CDM cosmology suggest different
explanations for those discrepancies. Thus for instance, according to 
\cite{Kogut}, the older
proposal of \cite{Matzner} in which  matter 
and CMB velocities could have started to differ after decoupling, 
predicts the existence 
of bulk motions over horizon scales.
Other hypothesis is that the dipole is not kinematic, 
but of cosmological origin, due to
an entropy gradient (isocurvature perturbation) on super-Hubble 
scales \cite{Paczynski,Langlois} or to  preinflationary
remnants in the density perturbations \cite{Turner}. 
In any case, according to
the previous discussion, the possibility that matter and radiation have different
rest frames is not observationally excluded.

However matter and radiation are not the dominant
components of the universe today. Recent observations of SNIa \cite{SN} combined with CMB 
anisotropies \cite{WMAP} suggest that the rate of expansion of the universe
is increasing, the
acceleration being driven by an unknown form of dark energy with a
relative  density $\Omega_\Lambda= 0.73\pm 0.04$ and equation
of state $w_\Lambda<-0.78$ (at the 95\% c.l.). (For the more 
recent three-year WMAP data \cite{WMAP3} $\Omega_\Lambda=0.72\pm 0.04$
and $w_\Lambda=-0.97^{+0.07}_{-0.09}$ (WMAP+SNLS) for a
flat universe).  Although 
the nature of dark energy is a complete mistery,
several models have been proposed in which dark energy appears 
either as a pure cosmological constant term in the Einstein equations; as
a perfect fluid with appropriate equation of state; 
as an extremely light scalar field running down the slope of a given
potential in the quintessence models \cite{quintessence}; or
as a scalar field with non-canonical kinetic term (k-essence) \cite{kessence}. 
In all those cases, dark energy
is completely decoupled from matter and radiation, its only 
effects being of gravitational  nature (see \cite{DE} and references 
therein).

The existence of a common rest frame is expected for strongly coupled fluids,
 as is indeed
the case for baryonic matter and radiation before recombination. 
However, this might not be true at the present epoch when matter,
radiation, and presumably dark energy are almost completely 
decoupled. 
In such a case, it makes sense to explore the 
possibility that the different components have different rest frames.

\section{Velocity perturbations}

Let us therefore consider  a cosmological scenario with three perfect fluids: 
radiation, matter and dark energy, whose equations of state read 
$p_\alpha=w_\alpha \rho_\alpha$ with  $\alpha=R,M,\Lambda$. For the sake
of generality, we will allow the dark energy equation of state to
have a smooth dependence on redshift $w_\Lambda(z)$. 
The energy-momentum tensor of each fluid will take the form: 
\begin{eqnarray}
(T^\mu_{\;\;\nu})_\alpha=(\rho_\alpha+p_\alpha)
 u^\mu_\alpha u_{\nu\alpha}- p_\alpha\delta^\mu_{\;\;\nu}
\end{eqnarray}
Since in this work we are only interested in the effects of fluids
motion on the CMB dipole, it is sufficient to take into account
 only the evolution of velocity
perturbations, i.e. we will not consider density or pressure perturbations.
In addition, since we will only concentrate on the dipole anisotropy, it is 
enough to consider the homogeneous part of the velocity fields. The 
presence of inhomogeneities will contribute to higher multipoles, 
(see for instance \cite{Gio}). 
Therefore, for this particular problem we can write:
\begin{eqnarray}
\rho_\alpha&=&\rho_\alpha(\eta),\nonumber \\ 
 p_\alpha&=&p_\alpha(\eta), \nonumber \\ 
 u^\mu_\alpha&=&\frac{1}{a}(1,v^i_\alpha(\eta))
\end{eqnarray} 
where $\eta$ is the conformal time.    
The effects of dark energy perturbations on 
higher multipoles have been considered in \cite{Moroi}, mainly 
in connection with the problem of the low CMB quadrupole.

 In the following we will assume that
$\vec v_\alpha^{\,2}\ll 1$ and we will work at first order 
in perturbation theory. Since the fluids only support
velocity perturbations, the form of the space-time metric will be given
by the most general vector-perturbed 
Friedmann-Robertson-Walker metric:
\begin{eqnarray}
ds^2=a^2(\eta)\left(d\eta^2+2S_i\,d\eta\, dx^i-(\delta_{ij}+2F_{i,j})
\,dx^i\,dx^j\right)
\label{metric}
\end{eqnarray}
Notice that we are assuming a spatially-flat
universe and accordingly for the unperturbed spatial metric we have
$g_{ij}=-a^2\delta_{ij}$. The vector perturbations $S_i$ and $F_i$ are
written in the notation of \cite{Mukhanov}. 
As commented above,  
for the dipole contribution we  consider fluid velocities 
 depending only on time, i.e. we will limit ourselves to the zero-mode 
equations, and therefore,  
we can take $S_i=S_i(\eta)$ and $F_{i,j}=0$. Accordingly,  
the total energy-momentum tensor reads:
\begin{eqnarray}
T^0_{\;\;0}&=&\sum_\alpha \rho_\alpha\nonumber \\
T^0_{\;\;i}&=&\sum_\alpha (\rho_\alpha+p_\alpha)(S_i-v_{i\alpha})\nonumber \\
T^i_{\;\;0}&=&\sum_\alpha (\rho_\alpha+p_\alpha)v^i_\alpha\nonumber \\
T^i_{\;\;j}&=&-\sum_\alpha p_\alpha\delta^i_{\;\;j}
\label{T}
\end{eqnarray}  
Notice that we are considering only the epoch after matter-radiation 
decoupling, assuming that dark energy is also decoupled and for that
reason we will ignore possible energy and momentum transfer
effects. 

We now calculate the linearized Einstein equations using
(\ref{metric}) and (\ref{T}).  The $(^0_{\;\;0})$ and $(^i_{\;\;j})$
components are trivial, whereas the $(^0_{\;\;i})$ and $(^i_{\;\;0})$
yield the condition:
\begin{eqnarray}
S^i=\frac{\sum_\alpha (\rho_\alpha+p_\alpha)v^i_\alpha}
{\sum_\alpha (\rho_\alpha+p_\alpha)}
\label{S}
\end{eqnarray}
On the other hand, the energy conservation equations are trivially
satisfied, whereas the total momentum conservation implies:
\begin{eqnarray}
\frac{d}{d\eta}
\left(a^4\sum_\alpha (\rho_\alpha+p_\alpha)(S^i-v^i_\alpha)\right)=0
\end{eqnarray}
which is compatible with (\ref{S}). In General Relativity the combination 
$(\rho_\alpha+p_\alpha)$ 
appearing in (\ref{S}) 
plays the role of inertial mass density of the corresponding fluid 
(see \cite{Schutz}), and 
accordingly $S^i$ can be understood as the 
{\it cosmic center of mass velocity}. Notice that a pure cosmological
constant has no inertial mass density. Since 
matter-radiation decoupling takes place in the matter dominated era, the
cosmic center of mass velocity is determined after decoupling 
by the motion of matter and dark energy, radiation playing essentially
no role.

The momentum conservation equation for each fluid:
\begin{eqnarray}
\frac{d}{d\eta}(a^4(\rho_\alpha+p_\alpha)(S^i-v^i_\alpha))=0
\label{vel}
\end{eqnarray}
 implies that
the corresponding velocity relative to the center of mass frame scales
as: $\vert \vec S-\vec v_\alpha\vert\propto a^{3w_\alpha-1}$, i.e.
it is constant in the case of radiation and scales as $a^{-1}$ for matter. 
In the case of dark energy the scaling properties will depend on the particular
model under consideration, as we will see below. Notice that for the 
zero modes, the equations above contain all the information about the
evolution of the velocity perturbations.

\section{Effects on the CMB dipole}

Once we know the form of the perturbed metric, we can calculate the
effect of fluids motion on photons propagating from the last scattering
surface using standard tools (see for instance \cite{Gio} and references
therein). 
The energy of a photon coming from direction 
$n^\mu=(1,n^i)$ with $\vec n^{\,2}=1$ as seen by an observer moving with
velocity $u^\mu=a^{-1}(1,v^i)$ is given by:
\begin{eqnarray}
{\cal E}=g_{\mu\nu}u^\mu P^\nu
\end{eqnarray} 
with
\begin{eqnarray}
P^\nu= \frac{E}{a}\left(n^\nu+\frac{d\delta x^\nu}{d\eta}\right)
\end{eqnarray}
where $E$ parametrizes the photon energy and the perturbed 
trajectory of the photon reads 
$x^\mu(\eta)=x_0^\mu(\eta)+\delta x^\mu$, with $x_0^\mu=n^\mu \eta$. 
 To first order in the perturbation, 
assuming that the observer velocity is of the same order as the metric
perturbation, we get:
\begin{eqnarray}
{\cal E}\simeq \frac{E}{a}\left(1+\frac{d\delta x^0}{d\eta}+\vec n\cdot
(\vec S- \vec v)\right)
\end{eqnarray}

In order to obtain $d\delta x^0/d\eta$, we solve the geodesics equations
to first order in the perturbations. In order to simplify the calculation, 
we notice that the geodesics corresponding to the $g_{\mu\nu}$ metric
with affine parameter $\tau$
are the same as those corresponding to the 
$\hat g_{\mu\nu}=a^{-2}g_{\mu\nu}$ metric, with parameter $\eta$
such that $d\tau=a^2d\eta$. In such a case the unperturbed 
$\hat g_{\mu\nu}$ is nothing but the Minkowski metric, and 
the 0-component of the geodesics equation reduces to:
\begin{eqnarray}
\frac{d^2\delta x^0}{d\eta^2}=0
\label{geodesics}
\end{eqnarray}
By defining $\hat{\cal E}=a{\cal E}$, the temperature fluctuation
generated by the Sachs-Wolfe effect in this particularly simple case reads:
\begin{eqnarray}
\left.\frac{\delta T}{T}\right\vert_{dipole}&=&
\frac{\hat{\cal E}_0-\hat{\cal E}_{dec}}{\hat{\cal E}_{dec}}\simeq
\left.\frac{d\delta x^0}{d\eta}\right\vert^0_{dec}+\vec n\cdot
(\vec S- \vec v)\vert^0_{dec}\nonumber \\
&\simeq&\vec n\cdot
(\vec S- \vec v)\vert^0_{dec}
\end{eqnarray}
where the indices $0$, $dec$ denote the present and decoupling times
respectively, and we have made use of 
 (\ref{geodesics}). 

At decoupling, the universe is matter dominated and it is a good 
approximation to neglect the contribution to $\vec S$
from dark
energy. Since baryons and radiation were coupled until
recombination, we take the 
velocity of matter $\vec v_M^{dec}$ to be  the same as 
that of radiation at that time $\vec v_R^{dec}$, and accordingly 
we have $\vec S_{dec}\simeq \vec v_M^{dec}\simeq \vec v_R^{dec}$.
Here we  are assuming for simplicity that baryonic and dark matter share
a common rest frame.
On the other hand, if we assume that the 
intrinsic density fluctuations in the last scattering
surface contribute only a small fraction to the CMB dipole,   
it is a good approximation to take the emitter velocity to 
be $\vec v_{dec}\simeq \vec v_M^{dec}$.

On the other hand, today the contribution of radiation to
the energy density is negligible and from (\ref{S}) we get:
\begin{eqnarray}
\vec S_0\simeq\frac{\Omega_M\vec v_M^0+(1+w_{\Lambda}^0)\Omega_{\Lambda}
\vec v_{\Lambda}^0}
{1+w_{\Lambda}^0\Omega_{\Lambda}}
\end{eqnarray}
so that we find:
\begin{eqnarray}
\left.\frac{\delta T}{T}\right\vert_{dipole}&\simeq&
\vec n\cdot(\vec S_0- \vec v_0) \label{dipole}\\
&\simeq&\vec n \cdot \frac{
\Omega_M(\vec v_M^0- \vec v_0)
+(1+w_{\Lambda}^0)\Omega_\Lambda(\vec v_\Lambda^0- \vec v_0)}
{1+w_{\Lambda}^0\Omega_{\Lambda}}\nonumber
\end{eqnarray}
where $w_\Lambda^0=w_\Lambda(0)$ is the present value of the 
dark energy equation of state and we have used $\Omega_M+\Omega_\Lambda=1$.

According to this result, the CMB dipole is due
to the relative velocity of the observer with respect to the present cosmic 
center of mass. In the particular case in which matter, radiation and
dark energy share a common rest frame, i.e. 
$\vec v_M^0=\vec v_R^0=\vec v_\Lambda^0$
then the previous result reduces to the usual expression for the dipole:
$\delta T/T\vert_{dipole}\simeq 
\vec n\cdot(\vec v_R^0- \vec v_0)$. 
However this needs not to be necessarily the case. Thus, in particular,
it is possible that an observer at rest with  radiation 
$\vec v_0=\vec v_R^0\neq\vec v_M^0 \neq \vec v_\Lambda^0$ can measure an
nonvanishing dipole according to (\ref{dipole}).

Although in  Standard Cosmology we expect  
$\vec v_R^0=\vec v_M^0$, provided matter 
and radiation shared a common rest frame until decoupling, 
we have seen that the existence of large scale bulk flows would suggest
this not to be the case.  There are indeed proposals in the 
literature in which matter and radiation
rest frames started to differ after recombination \cite{Matzner}. For that 
reason, in this work, we have allowed  for possible velocity differences 
today.

\section{Matter bulk flows and moving dark energy}

In the absence of dark energy or in the case in which it is in the 
form of a pure 
cosmological constant ($w_\Lambda=-1$), dark energy would not contribute 
to the center of mass motion. Moreover, today the radiation contribution
is negligible and accordingly the center of mass
rest frame would coincide with the matter rest frame.
As commented above, there are works  
in which matter and radiation
rest frames start to differ after decoupling, 
and such an offset has been claimed to generate bulk flows on
large scales. However our results imply 
that  the relative motion of matter and radiation
today could not 
explain the existence of  bulk flows on the largest scales, since the 
frame in which the dipole vanishes would coincide with the matter
rest frame.  Conversely, the 
existence of non-vanishing bulk flows would require the presence 
of moving dark energy with $w_\Lambda^0\neq -1$. 

Indeed, if moving dark energy is responsible for the 
existence of cosmic bulk flows  on very large scales, 
then the amplitude and direction of such flows would 
provide a direct measurement
of the relative velocity of matter and dark energy. The bulk flow $\vec V_b$ 
can be
understood as the average velocity of a given matter volume with respect
to an observer  who  measures a vanishing CMB dipole, i.e. 
$\vec V_b=\vec v_M^0-\vec v_0$. Such an 
observer has a  velocity which is given, according to (\ref{dipole}), by:
\begin{eqnarray}
\vec v_0\simeq \vec v_M^0+\frac{(1+w_\Lambda^0)\Omega_\Lambda}
{1+w_\Lambda^0\Omega_\Lambda}
(\vec v_\Lambda^0-\vec v_M^0)
\end{eqnarray}
so that 
\begin{eqnarray}
\vec v_M^0-\vec v_\Lambda^0 \simeq \frac{1+w_\Lambda^0\Omega_\Lambda}
{(1+w_\Lambda^0)\Omega_\Lambda} \vec V_b=P_\Lambda \vec V_b
\end{eqnarray}

Notice that curiously, according to these results, 
even if matter is at rest with respect to the CMB radiation, 
$\vec v_M^0=\vec v_R^0$, 
it would be possible to have a non-vanishing flow $\vec V_b\neq 0$, provided
dark energy is moving with respect to matter. 
The proportionality constant
$P_\Lambda$ depends on the present value of $w_\Lambda$.
Thus for instance, taking $\Omega_\Lambda=0.73$, we get $P_\Lambda=2.68$ for
$w_\Lambda^0=-0.78$ or $P_\Lambda=14$ for the central values
of the WMAP three-year data $\Omega_\Lambda=0.72$ and $w_\Lambda^0=-0.97$. 

\section{Discussion}

Notice that bulk  velocities of several hundred km s$^{-1}$ are
obtained in some recent large scale bulk flow measurements. 
Therefore, in order 
for moving dark energy to have measurable effects,   
the relative matter-dark energy
relative velocity today should be in the range $\gsim 10^2$ km s$^{-1}$. 
Since the nature of dark energy is still unknown, and there is 
not a generally accepted model, we can only limit ourselves to
show that those relative velocities are not a priori excluded, and
that, in fact, there is a wide class of models in which such values
can be obtained in a natural way.

As implied by (\ref{vel}), the relative velocity of dark energy
with respect to the center of mass frame scales as 
$\vert \vec S-\vec v_\Lambda\vert\propto a^{3w_\Lambda(z)-1}$. 
Thus, in models  with constant equation of state,  the relative velocity
 decreases faster than $a^{-3.3}$, since as commented before $w_\Lambda<-0.78$, 
 so that any initial
relative velocity is rapidly damped away. 
However there is another  class 
of  models in which the dark energy equation of
state exhibits {\it scaling} behavior, i.e., 
it mimics that of the 
dominant component of the universe during most of the cosmological
evolution. Such models have in addition the interesting
property of being able to  alleviate the fine tuning problem of
 models  with constant equation of state. Examples of
scaling models  include
quintessence \cite{quintessence}, k-essence \cite{kessence} 
and  other phenomenological proposals (see \cite{scaling} and references
therein).  
In such models  
$w_\Lambda(z)\simeq 1/3$ for $z\gg z_s$,  $w_\Lambda(z)\simeq 0$       
for $z_b\ll z\ll z_s$ and $w_\Lambda(z)\simeq -1$ for $z\ll z_b$ 
($z_s\sim \Od(10^3)$ and $z_b\sim \Od(1)$ being typical values \cite{scaling}).
This means that, according to the above scaling 
behavior, no damping of the relative velocity would have taken
place until $z\sim z_s$. Then a mild damping $\sim a^{-1}$  
would have occurred  for $z_b\lsim z\lsim z_s$, and only very recently 
($z\lsim z_b$), the damping
would have been stronger  $\sim a^{-4}$, i.e. we expect typical 
total damping factors 
today around $10^{-3} - 10^{-4}$ 
(depending on the exact transition redshifts).  If the nature of dark energy
is really independent of the rest of components of the universe, and it 
has been always decoupled from them, then the corresponding initial dark energy
bulk velocity should be considered as a free cosmological parameter. 
Accordingly the present relative velocity would be  comparable to the amplitudes 
of the observed CMB dipole and  bulk flows measurements for initial values around
$\vert \vec S-\vec v_\Lambda\vert \lsim 1$. 

A detailed analysis of the 
different dark energy models will be presented elsewhere \cite{Maroto}.
 In any case, regardless the particular mechanism responsible for
the motion of dark energy,  it is interesting to note, according to
the previous discussion,  that a 
better determination
of the existence of matter flows on very large  scales  could  shed light 
on the nature of dark energy.

Finally, we would  like to mention that 
the metric anisotropies created by the relative motion of
fluids \cite{2fluids} could also have  observable effects on photons and other 
types of particles propagating from sources located at cosmological
distances \cite{Lorentz}. Such effects could offer independent evidence
of the motion of dark energy \cite{Maroto}.

\vspace{.2cm}

{\em Acknowledgments:} 
This work has been partially supported by DGICYT (Spain) under project numbers
FPA 2004-02602 and FPA 2005-02327

\end{document}